\documentclass[preprint,amsmath,showpacs,amssymb,prb,a4paper]{revtex4}
\usepackage{graphicx}
\usepackage{bm}

\def\unitev{\,{\rm eV}}
\def\unitmev{\,{\rm meV}}
\def\unitnm{\,{\rm nm}}
\def\unitwn{\,{\rm cm}^{-1}}
\def\unitps{\,{\rm ps}}
\def\unitfs{\,{\rm fs}}

\def\mod{\textrm{mod}}
\def\exph{M_{\rm ex-ph}}
\def\exop{M_{\rm ex-op}}
\def\elph{M_{\rm el-ph}}

\def\imag{{\rm i}}
\def\euler{{\rm e}}
\def\diff{{\rm d}}

\begin{document}

\title{Excitonic effects on coherent phonon dynamics in single wall
  carbon nanotubes}

\author{A.~R.~T.~Nugraha$^1$, E.~Rosenthal$^{1,2}$, E.~H.~Hasdeo$^1$,
  G.~D. Sanders$^3$, C.~J.~Stanton$^3$, M.~S.~Dresselhaus$^{4}$,
  R.~Saito$^1$}

\affiliation{ $^1$Department of Physics, Tohoku University, Sendai
  980-8578, Japan\\
  $^2$Department of Physics, University of Pennsylvania,
  Philadelphia, Pennsylvania 19104, USA \\
  $^3$Department of Physics, University of Florida, Box 118440,
  Gainesville, Florida 32611-8440, USA\\
  $^4$Department of Physics, Massachusetts Institute of Technology,
  Cambridge, MA 02139-4307, USA}

\date{\today}

\begin{abstract}
  We discuss how excitons can affect the generation of coherent radial
  breathing modes in the ultrafast spectroscopy of single wall carbon
  nanotubes.  Photoexcited excitons can be localized spatially and
  give rise to a spatially distributed driving force in real space
  which involves many phonon wavevectors of the exciton-phonon
  interaction.  The equation of motion for the coherent phonons is
  modeled phenomenologically by the Klein-Gordon equation, which we
  solve for the oscillation amplitudes as a function of space and
  time.  By averaging the calculated amplitudes per nanotube length,
  we obtain time-dependent coherent phonon amplitudes that resemble
  the homogeneous oscillations that are observed in some pump-probe
  experiments.  We interpret this result to mean that the experiments
  are only able to see a spatial average of coherent phonon
  oscillations over the wavelength of light in carbon nanotubes and
  the microscopic details are averaged out.  Our interpretation is
  justified by calculating the time-dependent absorption spectra
  resulting from the macroscopic atomic displacements induced by the
  coherent phonon oscillations.  The calculated coherent phonon
  spectra including excitonic effects show the experimentally observed
  symmetric peaks at the nanotube transition energies, in contrast to
  the asymmetric peaks that would be obtained if excitonic effects
  were not included.
\end{abstract}

\date{\today}
\maketitle

\section{Introduction}
Single wall carbon nanotubes (SWNTs) have been an important material
for providing a one-dimensional (1D) model system to study the
dynamics and interactions of electrons and phonons.  These properties
are known to be very sensitive to the SWNT geometrical structure,
characterized by the chiral indices $(n,m)$.~\cite{c471} With rapid
advances in ultrafast pump-probe spectroscopy, it has recently been
possible to observe lattice vibrations of SWNTs in real time by
pump-probe measurements, corresponding to coherent phonon
oscillations.~\cite{gambetta06-cp,lim06-cpexp,kato08-cpaligned,
  kim09-cpprl,makino09-cpdoping} Femtosecond laser pump pulses applied
to a SWNT induce photo-excited electron-hole pairs bound by the
Coulomb interaction, called
excitons.~\cite{gambetta06-cp,kilina07-adv} Shortly after the excitons
relax to the lowest exciton states ($\approx 10\unitfs$), the SWNT
starts to vibrate coherently by exciton-phonon interactions because
the driving forces of the coherent vibration by excitons act at the
same time.

The coherent phonon motions can be observed as oscillations of either
the differential transmittance or the reflectivity of the probed light
as a function of delay time between the pump and probe pulses.  By
taking a Fourier transformation of the oscillations with respect to
time, we obtain the coherent phonon spectra as a function of the
phonon frequencies.  Several peaks found in the coherent phonon
spectra correspond to certain optically active phonon modes.  Typical
SWNT phonon modes observed from the coherent phonon spectra are
similar to those found in the Raman spectra because the exciton-phonon
interactions are responsible for both coherent phonon excitations and
Raman spectroscopy.  However, unlike Raman spectroscopy, ultrafast
spectroscopy techniques allow us to directly measure the phonon
dynamics, including phase information, in the time
domain.~\cite{gambetta06-cp,lim06-cpexp,kim09-cpprl}

One of most commonly observed coherent phonon modes in SWNTs is the
radial breathing mode (RBM), in which the tube diameter vibrates by
initially expanding or contracting depending on the tube types
$(\mod(n-m,3) = 0,1,2)$ and excitation energies.~\cite{kim09-cpprl}
Previously we have developed a microscopic theory for the
type-dependent generation of coherent RBM phonons in SWNTs within an
extended tight binding model and effective mass theory for
electron-phonon interactions.~\cite{sanders09-cp, nugraha11-cp} This
model did not take into account the excitonic interaction between the
photoexcited electrons and holes.  We found that such initial
expansion and contraction of the SWNT diameter originates from the
wavevector-dependent electron-phonon interactions in SWNTs.  Although
the coherent phonon generation mechanism neglecting exciton effects
considered in previous studies could describe some main features of
the coherent phonons in SWNTs, it predicted an asymmetric line shape
in contrast to the experimentally observed symmetric line shape.  This
discrepancy indicates that the presence of excitons in SWNTs should be
important microscopically.~\cite{ando97-exc, spa04-exc, wang05-exp,
  jiang07-exc}

Excitons should have at least four important effects on the generation
and detection of coherent phonons in SWNTs: (1)~the optical
transitions will be shifted to lower energy owing to the Coloumb
interaction between the photoexcited electron-hole
pair,~\cite{ando97-exc} (2)~the strength of the optical transitions
will be enhanced since the excitonic wavefunctions have larger optical
matrix elements resulting from the localized exciton
wavefunctions,~\cite{jiang07-exphop} (3)~the phonon interaction matrix
elements may also change because the electron-phonon and hole-phonon
matrix elements now become exciton-phonon matrix
elements,~\cite{jiang07-exphop} and (4)~in SWNTs, the excitons can
become localized along the tube with a typical exciton size of about
$1\unitnm$.~\cite{carsten10-exclocal} This will change which phonon
modes can couple to the photogenerated excitons.  Excitons are known
to have localized wavefunctions in both real and reciprocal
space,~\cite{jiang07-exc} and this should modify the electron-phonon
picture of the coherent phonon generation.  Due to the localized
exciton wavefunctions, the driving force of a coherent phonon is
expected to be a Gaussian-like driving force in real space for each
localized exciton, whose width is about $1\unitnm$, instead of a
constant force considered in the previous
works.~\cite{sanders09-cp,nugraha11-cp} The localized force can be
obtained only if we consider the coupling of excitons and phonons.

The interaction between excitons and coherent phonons will involve
many phonon wavevectors for making localized vibrations and many
electron (and hole) wavevectors for describing these excitons.  By
applying strong pump light to the SWNTs, many excitons are generated
and the average distances between two nearest excitons are estimated
to be about $20\unitnm$.~\cite{kamm07-biexciton, matsuda08-exciton}
This indicates that the driving force for coherent phonon generation
can be approximated by many Gaussians, each of which originates from
an exciton and are separated by the distance between two excitons.
Using such a driving force model also implies that the coherent phonon
amplitudes are inhomogeneous along the nanotube axis.  However, since
the wavelength of light ($\sim{500\unitnm}$) is much larger than the
spatial modification of the RBM amplitudes, the laser light can only
probe the average of the coherent vibrations.

To simulate the exciton effects using coherent phonon spectroscopy, we
model the coherent RBM phonon amplitude $Q(z,t)$ as a function of
space and time using the Klein-Gordon equation that will be shown to
explain the dispersive wave properties.  The driving forces are
localized almost periodically, and therefore the calculated coherent
phonon amplitudes of the RBM are no longer constant along the tube
axis.  However, by taking an average over the tube length for the
calculated coherent phonon amplitudes, we find that the average
amplitude fits the oscillations as a function of time observed in the
experiments.  In order to compare our theory directly with
experiments, in which the change of the transmittance $(\Delta T / T)$
or reflectivity $(\Delta R / R)$ is measured, we calculate the
time-dependent absorption spectra for macroscopic atomic displacements
induced by the coherent phonon oscillations $Q(z,t)$.  The symmetric
line shape found in the calculated spectra is also consistent with the
experimental observations.

This paper is organized as follows.  In Section II, we give the
phenomenological model for the generation of coherent RBM phonons,
which is expressed by the Klein-Gordon equation.  The Klein-Gordon
equation is able to explain the propagation of the coherent RBM
phonons induced by excitons because it gives the RBM phonon
dispersion.  In Section III, we present the main results and discuss
how the inhomogeneous coherent amplitudes obtained from solving the
Klein-Gordon equation can lead to the observed homogeneous
time-dependent absorption spectra.  Finally, we give conclusions in
Section IV.

\section{Coherent phonon model}
\label{sec:cpa}

In the conventional model for the coherent phonon generation mechanism
in semiconductor systems, the phonon modes that are typically excited
are the ones with phonon wavevector $q = 0$.  The coherent phonon
amplitudes $Q_{\rm c}(t)$ satisfy a driven oscillator
equation,~\cite{stanton94-cpmethod,merlin97-cp}
\begin{equation}
  \frac{\partial^2 Q_{\rm c}(t)}{\partial t^2} +
  \omega_{\rm 0}^2 Q_{\rm c}(t) = S_{\rm c}(t) ,
\label{eq:drivenoscillator}
\end{equation}
where $\omega_0$ is the phonon frequency at $q = 0$ and $S_{\rm c}(t)$
is a driving force that depends on the physical properties of a
specific material.  In the case of a SWNT, without considering the
excitonic effects, $S_{\rm c}(t)$ is given
by~\cite{sanders09-cp,nugraha11-cp}
\begin{equation}
  S_{\rm c}(t) = -\frac{2 \ \omega_0}{\hbar} \sum_{\mu k} \elph^\mu (k) \delta
  f^\mu (k,t),
\label{eq:convdrivenforce}
\end{equation}
where $\elph^\mu (k)$ is the electron-phonon matrix element for the
$\mu$-th cutting line (one-dimensional Brillouin zone of a SWNT) as a
function of the one-dimensional electron wavevector $k$ and is
calculated for each phonon mode at $q = 0$.  The distribution function
$\delta f^\mu$ of photo-excited carriers generated by a laser pulse
pumping at the $E_{ii}$ transition energy is obtained by solving a
Boltzmann equation for the photogeneration
process.~\cite{sanders09-cp}

We can see in Eqs.~(\ref{eq:drivenoscillator}) and
(\ref{eq:convdrivenforce}) that $Q_{\rm c}(t)$ and $S_{\rm c}(t)$ have
a time dependence only and no spatial dependence when we consider
electron-photon (or hole-photon) and electron-phonon (or hole-phonon)
interactions, i.e. we ignored the excitonic interaction between the
photoexcited electrons and holes.  We now extend this model by
considering that the exciton effects (exciton-photon and
exciton-phonon interactions) give a spatial dependence to the coherent
phonon amplitude and to the driving force, which we denote as $Q(z,t)$
and $S(z,t)$, respectively.  Here $z$ is the position along the
nanotube axis.  To describe the coherent phonon amplitude $Q(z,t)$, we
propose using the Klein-Gordon equation,
\begin{equation}
\frac{\partial^2 Q(z,t)}{\partial t^2} -
c^2 \frac{\partial^2 Q(z,t)}{\partial z^2} = S(z,t)
-\kappa Q(z,t)
\label{eq:osc}
\end{equation}
where $c$ and $\kappa$ are the propagation speed and dispersion
parameter depending on the SWNT structure, respectively.  The
Klein-Gordon equation is solved subject to the two initial conditions
$Q(z,0) = 0$ and $\dot Q(z,0) = 0$.  The exciton-induced driving force
$S(z,t)$ is given by
\begin{equation}
  S(z,t) = - \frac{2}{\hbar} \sum_{\mu, k,q} \omega_q \exph^\mu (k,q) \delta
  f^\mu (k,t) \euler^{\imag q z}, \label{eq:force}
\end{equation}
where $\exph^\mu (k,q)$ is the exciton-phonon matrix element on the
$\mu$-th cutting line as a function of the exciton wavevector $k$ and
phonon wavevector $q$.  By using the driving force expression of
Eq.~\eqref{eq:force}, the amplitude $Q(z,t)$ is dimensionless because
the dimension of $S(z,t)$ is the inverse of time square (instead of
length per inverse of time square).  Here the actual coherent phonon
amplitudes with units of length can be obtained by multiplying
$Q(z,t)$ with the zero-point phonon amplitude $Q_0 = \sqrt{\hbar / 2
  M_{\rm c}\omega_0}$, where $M_{\rm c}$ is the total mass of the
carbon atoms in the nanotube unit cell.

\begin{figure}[tb]
  \centering \includegraphics[clip,width=8cm]{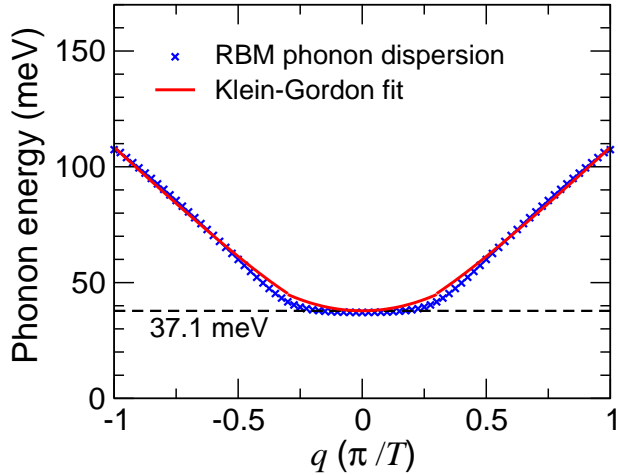}
  \caption{\label{fig:RBMdisp} (Color online) RBM phonon dispersion of
    a (11,0) nanotube.  Theoretical data are represented by cross
    symbols, which are calculated using a force constant model as in
    Refs.~\onlinecite{sanders09-cp} and \onlinecite{jishi93-phonon}.
    The solid line shows the fitted RBM dispersion using the
    Klein-Gordon dispersion relation in Eq.~\eqref{eq:quaddisp}.  The
    phonon energy, $\hbar \omega$, is plotted as a function of $q$ in
    units of $\pi / T$.  Here $T = 0.431\unitnm$ is the unit cell
    length of the $(11,0)$ tube. }
\end{figure}

The reason why we adopt the Klein-Gordon equation to explain the
exciton-induced coherent phonon generation in SWNTs is based on a
phenomenological consideration.  We generally expect that the coherent
RBM phonons are propagating dispersively along the nanotube axis.
Integrating $Q(z,t)$ and $S(z,t)$ over $z$ should give $Q_{\rm c}(t)$
and $S_{\rm c}(t)$ in Eq.~\eqref{eq:drivenoscillator} which describes
the homogeneous vibration observed in experiments.  Parameters $c$ and
$\kappa$ in the Klein-Gordon equation can then be obtained from the
RBM phonon dispersion, which gives positive $c$ and $\kappa$ values.
To obtain this relationship, we consider the Klein-Gordon
equation~\eqref{eq:osc} with $S(z,t) = 0$ and take a Fourier transform
defined by
\begin{equation} 
  \tilde Q(q, \omega) =
  \int\limits_{-\infty}^\infty \int\limits_{-\infty}^\infty Q(z,t)
  \euler^{\imag(qz - \omega t)} \diff{z}\diff{t}, 
\end{equation}
to obtain
\begin{equation}
  -\omega^2 \tilde{Q} + c^2 q^2 \tilde{Q} = -\kappa \tilde{Q}.
  \label{eq:ft}
\end{equation}
From Eq.~\eqref{eq:ft} we have a dispersion relation for the
Klein-Gordon equation,
\begin{equation}
  -\omega^2 + c^2 q^2  = -\kappa.
  \label{eq:disp}
\end{equation}
The physical solution of Eq~\eqref{eq:disp} for $\omega > 0$ is
\begin{equation}
  \omega (q) = \sqrt{c^2 q^2 + \kappa}.
\label{eq:quaddisp}
\end{equation}
We can then fit the wave dispersion to the RBM phonon dispersion which
is already available by force constant or first-principle
models.~\cite{jishi93-phonon,maul02-phonon,dubay03-phonon} We are
particularly interested in the region of $q \ll \pi/T$ ($T$ is the
unit cell length of a SWNT~\cite{c471}) because this is the typical
size over which an exciton in reciprocal space interacts with a
phonon.~\cite{jiang07-exc,jiang07-exphop} Fitting the RBM phonon
dispersion to Eq.~(\ref{eq:quaddisp}) thus gives the values of both
$c$ and $\kappa$ to be used in the Klein-Gordon equation.  As for the
phonon dispersion shown in Fig.~\ref{fig:RBMdisp}, which here is
calculated for a $(11,0)$ tube, we obtain $c = 2.545\unitnm/\unitps$
and $\kappa = 3147.22 \unitps^{-2}$.  Hereafter, we will consider the
$(11,0)$ tube as a representative example for the simulation.

\begin{figure}[tb]
  \centering \includegraphics[clip,width=8cm]{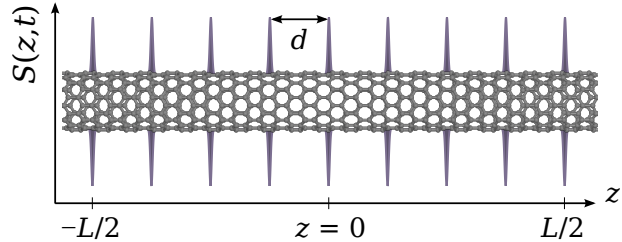}
  \caption{\label{fig:excitonforce} (Color online) Schematic
    illustration of the driving force $S(z,t)$ created by excitons
    which align along the nanotube axis.  In general, the excitons can
    be distributed randomly with an average separation between two
    excitons denoted by $d$.  The force $S(z,t)$ is symmetric in the
    circumferential direction.}
\end{figure}

To simulate the coherent phonon dynamics, we can further simplify the
driving force in Eq.~(\ref{eq:force}), which contains the
exciton-phonon matrix element, by assuming that the spatial shape of
the driving force follows that of the exciton wavefunction. This is
because the exciton-phonon matrix element is the electron-phonon
matrix element weighted by the exciton wavefunction coefficients. The
spatial shape of the exciton wavefunction can be fitted to a Gaussian
with a certain full width at half maximum, $\sigma_z$, that also
determines the exciton size. The exciton wavefunction with the
corresponding exciton energy dispersion can be obtained by solving the
Bethe-Salpeter equation.~\cite{jiang07-exc,jiang07-exphop}

Furthermore, we consider that the Gaussian force appears approximately
every $15-30\unitnm$ along the tube axis depending on the photoexcited
carrier density. For example, by solving for the photo-excited
distribution $\delta f$ using the method described in
Ref.~\onlinecite{sanders09-cp}, we estimate an exciton density for a
$(11,0)$ tube at an excitonic transition energy $E_{22} = 1.78\unitev$
which is about $5.6 \times 10^{-2}\unitnm^{-1}$.  This exciton density
corresponds to the average spatial separation between two excitons of
about $18\unitnm$.  In this case, we neglect the exciton
center-of-mass motion that involves the exciton-exciton interaction,
such as would be important for exciton diffusion and the Auger
effect,~\cite{kamm07-biexciton, matsuda08-exciton,konabe09-auger}
which could be considered in a future work.

Before the excitons interact with each other, the optically excited
exciton does not have the center-of-mass momentum because of the
energy-momentum conservation, and thus we need some more additional
time (sub-picoseconds) after the excitation to obtain the finite
diffusion constant which affects the coherent phonon dynamics.  In a
micelle-encapculated nanotube sample, excitons typically diffuse by
about $2 \unitnm$ (every $1\unitps$),~\cite{xie12-diff} while the
average separation between two excitons is one order of magnitude
larger.  Although in a pristine nanotube sample the excitons can
diffuse up to the same order as the average separation between two
excitons,~\cite{maru10-diffusion} the exciton diffusion mostly
contributes to the decay of the exciton life
time.~\cite{hagen05-decay} Also, the Auger rate is on the order of
$0.1\unitps^{-1}$, which corresponds to the ionization or
recombination times of excitons of about $10\unitps$,~\cite{onoda11}
whereas the time needed for generating coherent phonons in our case is
as early as hundred femtoseconds (the phonon period) and the time
scale for considering the coherent phonon dynamics is less than
$5\unitps$.  The Auger effect is then important in later time when any
two excitons can collide and disappear.  If the two excitons survive,
the coherent phonon amplitude may be given by a linear combination of
amplitudes induced by each exciton.  However, we did not consider such
situations for simplicity.  Therefore, in the present study, the total
driving force for the coherent phonon dynamics can be defined as a
summation of contributions from each Gaussian generated from an
exciton.

Each Gaussian function centered at the exciton position $z_i$, which
is distributed along the tube axis, is expressed as
\begin{equation}
  S_i(z,t) = A_g e^{-(z - z_i)^2/2 \sigma_z^2} \theta(t),
\label{eq:simforce}
\end{equation}
where $ \theta (t)$ is the Heaviside step function, $A_g$ is the force
magnitude obtained from the product of the exciton-phonon interaction
and the related factors in Eq.~\eqref{eq:force}, and $\sigma_z$ is the
width of the exciton-phonon matrix element for a given $(n,m)$ SWNT.
A typical value of $\sigma_z$ is related to the exciton size in real
space ($\sim 1\unitnm$).  The exciton wavefunctions, exciton energies,
exciton-photon and exciton-phonon matrix elements are all calculated
by solving the Bethe-Salpeter equation within the extended
tight-binding method as developed by Jiang {\it et
  al}.~\cite{jiang07-exc,jiang07-exphop} The force magnitude thus
obtained is on the order of $10^3\unitps^{-2}$.  For the lowest
$E_{22}$ exciton state of the $(11,0)$ tube, we obtain $\sigma_z =
0.9\unitnm$ and $A_g = 4.82 \times 10^{3} \unitps^{-2}$.  The total
driving force used in solving Eq.~(\ref{eq:osc}) is a summation of
Gaussian forces in terms of Eq.~(\ref{eq:simforce}),
\begin{equation}
  S(z,t) = \sum_{i=1}^N S_i(z,t) ,
\label{eq:totforce}
\end{equation}
where $N$ is the number of excitons (and thus the number of Gaussian
forces) in a SWNT.  In Fig.~\ref{fig:excitonforce}, we show a
schematic diagram of a typical model for our simulation.  The driving
force $S(z,t)$ has an axial symmetry and is aligned along the nanotube
axis with a separation distance of $d$.  To avoid any motions of the
center of mass, the general force $S({\bm r},t)$ should also satisfy a
sum rule,
\begin{equation}
  \displaystyle
  \int\limits_{-\infty}^\infty S(\bm r, t) \diff{\bm r} = 0,
\end{equation}
which is automatically satisfied for $S(z,t)$ in
Eq.~(\ref{eq:totforce}) because of the axial symmetry of the model, as
can also be understood from Fig~\ref{fig:excitonforce}.  In the
present calculation, we fix $d = 18\unitnm$, and there are $N=9$
narrow Gaussian forces arranged periodically (thus $L = 144\unitnm$).
The RBM phonon energy near $q = 0$ is $37.1~\unitmev$, corresponding
to a frequency $\omega = 297\unitwn$ and a vibration period $\tau =
0.112~\unitps$.

It should be noted that the specific details of the spatial
arrangement of the localized excitons are also mainly determined by
the exciton-exciton interaction.~\cite{kamm07-biexciton,
  matsuda08-exciton,konabe09-auger} However, we can simply take into
account the main point resulting from these exciton-exciton
interactions that the excitons will be stabilized and will arrange
themselves in a certain spatial configuration.  In general, excitons
do not need to be arranged periodically and can be distributed
randomly along the tube axis.  Here we use a specific exciton
configuration as a representative example that corresponds to a
slightly random configuration of excitons.  Interestingly, it will be
justified in the next section that even if the excitons are
distributed very randomly along the tube axis, the coherent phonon
amplitudes at each exciton site are not affected as far as two (or
more) excitons are not located at the same position.

\begin{figure}[tb]
  \centering \includegraphics[clip,width=7cm]{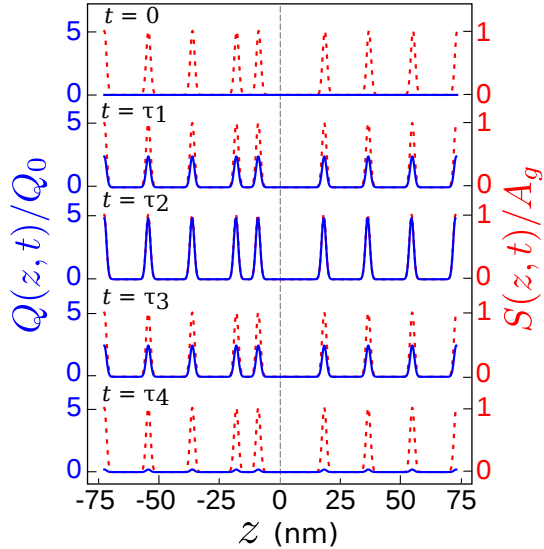}
  \caption{\label{fig:Qzt} (Color online) Time evolution of coherent
    phonon amplitudes in a $(11,0)$ nanotube for a slightly random
    distribution of excitons with an average separation $d=18\unitnm$
    and with the center force shifted by $9 \unitnm$.  Solid lines
    show snapshots of $Q(z,t)$ as a function of $z$ (position along
    the tube axis) for several different $t$ values with a time
    sequence $\tau_j = j\tau / 4$, where $\tau = 0.112\unitps$ is the
    fundamental period.  $Q(z,t)$ is plotted in terms of $Q_0 =
    2.59\times 10^{-3}\unitnm$.  Dotted lines show the force $S(z,t)$
    for comparison. }
\end{figure}

\section{Results and Discussion}

In Fig.~\ref{fig:Qzt}, we plot the coherent RBM phonon amplitudes
$Q(z,t)$ for a $(11,0)$ nanotube pumped at its $E_{22}$ transition
energy, in which a snapshot is taken for $t = 0$ to $\tau_4$, where
$\tau_j = j\tau / 4$.  We consider a slightly random configuration of
excitons with an average distance between two excitons $d = 18\unitnm$
and then we shift one of the excitons at the center of the tube axis
by $9\unitnm$.  The calculation is done numerically by solving for
$Q(z,t)$ from Eq.~(\ref{eq:osc}) with periodic boundary conditions at
$\pm L/2$.  We can observe some periodic peaks corresponding to each
localized force and these peaks also do not move as a function of
time.  One might then ask whether or not such exciton effects
correctly describe the coherent phonon oscillations in SWNTs.  This
can be answered by considering the average of the inhomogeneous
$Q(z,t)$ per nanotube length.

\begin{figure}[tb]
  \centering \includegraphics[clip,width=7cm]{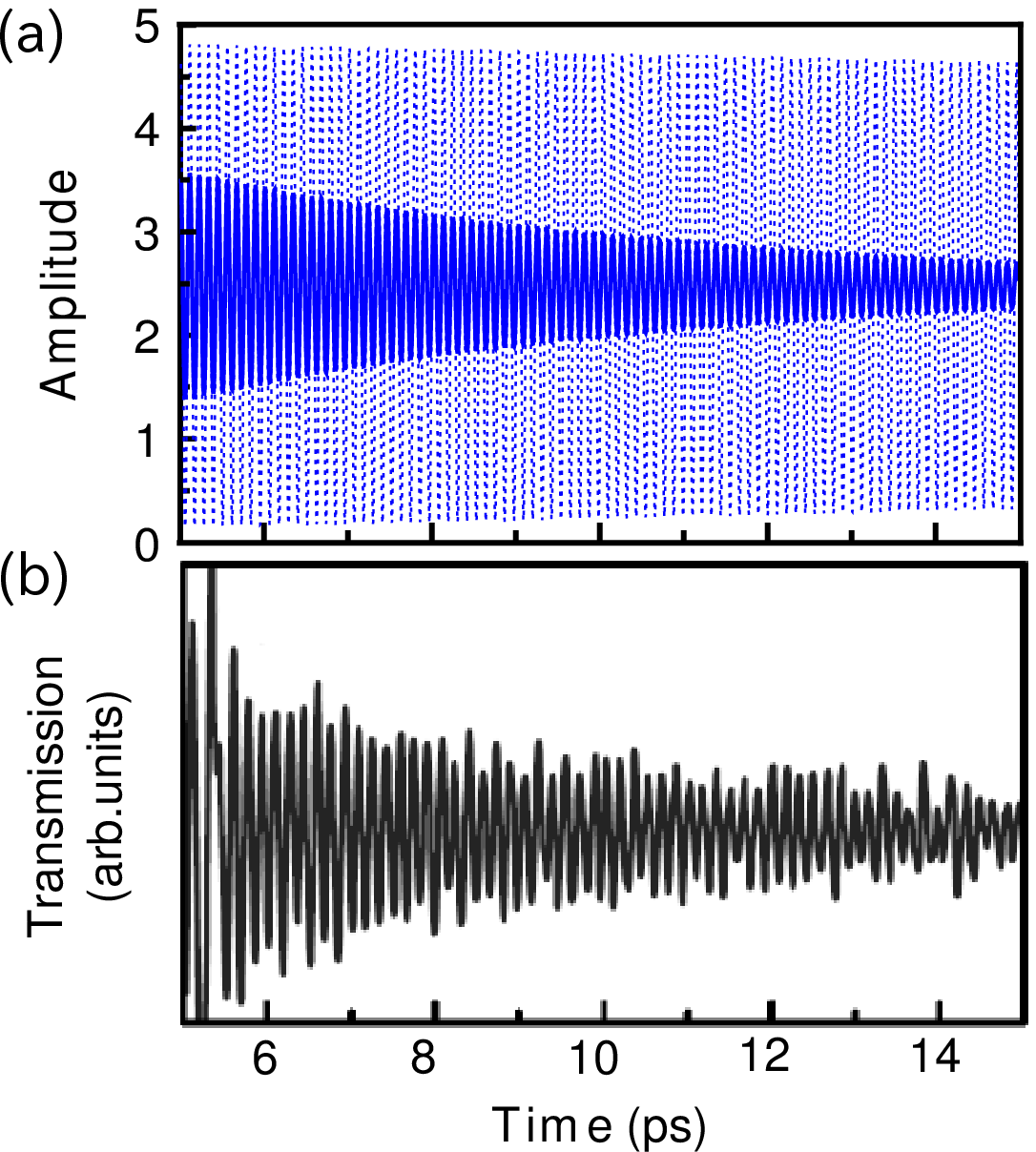}
  \caption{\label{fig:Aintegrated} (Color online) (a) Average of
    coherent phonon amplitudes per length, $A(t)$, plotted as a
    function of time for a $(11,0)$ nanotube ($\tau = 0.112\unitps$)
    and shown in units of $Q_0 = 2.59\times 10^{-3}\unitnm$.  The
    dotted line represents the average amplitude for the force
    distribution shown in Fig.~\ref{fig:Qzt}.  The solid line
    represents the average amplitude if a decay constant $0.2\ {\rm
      ps^{-1}}$ is taken into account.  (b) An example of the
    transmission oscillation data available for a $(13,3)$ tube
    measured in a pump-probe experiment with $\tau = 0.162\unitps$
    (reproduced from Ref.~\onlinecite{kim09-cpprl}).  The average
    coherent phonon amplitude shown in (a) resembles the oscillating
    feature of the experimental transmission shown in (b).}
\end{figure}

\begin{figure*}
  \centering
  \includegraphics[clip,width=16cm]{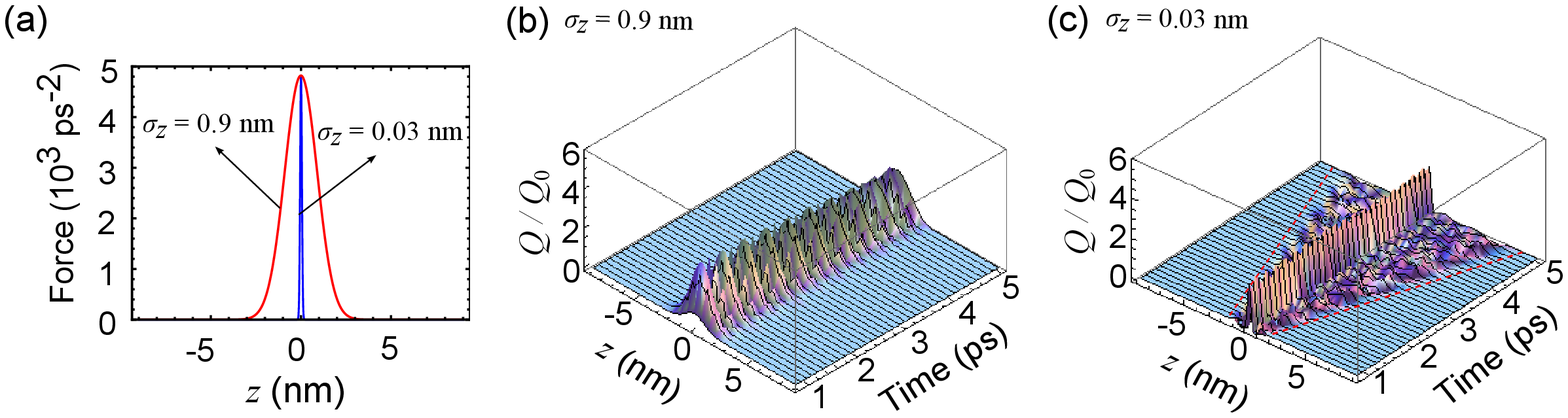}
  \caption{\label{fig:gaussdis} (Color online) (a) Driving forces with
    two different parameters $\sigma_z (=0.9\unitnm) > \sigma_{zc}$
    and $\sigma_z (=0.03\unitnm) < \sigma_{zc}$, which give (b) only
    localized and (c) both localized and propagating wave components,
    respectively.  For the $(11,0)$ tube in this simulation, we have
    $\sigma_{zc} = 0.045\unitnm$.  The propagating wave components in
    (c) travel with a speed of $1.68\unitnm/\unitps$.}
\end{figure*}

To clarify that our model can describe homogeneous coherent RBM phonon
oscillations that are observed in
experiments,~\cite{lim06-cpexp,kim09-cpprl} we define an average of
$Q(z,t)$ as follows
\begin{equation}
  A(t) = \frac{1}{L}\int_{L} Q(z,t) \diff z .
\label{eq:At}
\end{equation}
In Fig.~\ref{fig:Aintegrated}(a), we plot $A(t)$ for the $(11,0)$ tube
considered above.  We also include a decay constant $0.2\ {\rm
  ps^{-1}}$ to resemble the experimental results.~\cite{kim09-cpprl}
Interestingly, now the coherent phonon amplitudes, which have been
averaged before, could fit the experimental shape of the homogeneous
transmission oscillation in Figs.~\ref{fig:Aintegrated}(b).  We then
interpret that such an experiment cannot observe the nanoscopic
vibration of the exciton effects on the coherent phonon amplitudes,
but it can only observe the averaged amplitudes.  Moreover, the
definition~(\ref{eq:At}) is important mathematically to describe the
homogeneous coherent phonon amplitudes in experiments if we are able
to recover Eq.~(\ref{eq:drivenoscillator}) from the Klein-Gordon
equation~(\ref{eq:osc}).  Indeed, by integrating both left and right
sides of Eq.~(\ref{eq:osc}),
\begin{align}
  \int_L Q_{tt} \diff z - \int_L c^2 Q_{zz} \diff z = - \int_L \kappa
  Q \diff z + \int_L S \diff z, \notag
\end{align}
and using $\int_L Q_{tt} \diff z = A_{tt}$, $\int_L \kappa Q \diff z =
\kappa A$, $\int_L Q_{zz} \diff z = 0$, we can obtain
\begin{equation}
  A_{tt} + \kappa A(t) = S(z),
\end{equation}
which is nothing but the driven oscillator model in
Eq.~(\ref{eq:drivenoscillator}).

It is important to note that we have assumed a certain configuration
of excitons as a function of $z$.  However, excitons in nature might
not be uniformly spaced and any exciton distributions with random
spacing can be possible.  Nevertheless, we expect that our result for
the average amplitude $A(t)$ in Fig.~\ref{fig:Aintegrated} is
approximately constant regardless of the exciton spacing, as far as
the average exciton density remains the same.  This can be
rationalized by considering a trial solution of the Klein-Gordon
equation,
\begin{equation}
Q(z,t) = \euler^{-\lambda z}\euler^{i(q z-\omega t)},
\label{eq:trialsol}
\end{equation}
which comprises a travelling wave and a decay term with parameter
$\lambda$ to be determined.  By substituting Eq.~\eqref{eq:trialsol}
into Eq.~\eqref{eq:osc} and setting $S(z,t)=0$, we obtain
\begin{equation}
  \label{eq:acon}
  \lambda= iq \pm \sqrt{\frac{\kappa}{c^2}-q^2},
\end{equation}
where we have assumed $\omega = q c $ and the sign $\pm$ is determined
for the $\pm z$ region.  Depending on whether the value of
$\sqrt{\kappa/c^2 - q^2}$ is real or pure imaginary, respectively, we
can get a spatially localized or propagating solution of $Q(z,t)$.  In
the presence of a force, we can solve Eq.~\eqref{eq:osc} using the
Green's function method for a single Gaussian force $S(z,t)= A_g
\euler^{-z^2/2\sigma_z^2} \theta(t)$.  The solution for $Q(z,t)$ in
the region $-L/2 < z < L/2$ with a boundary condition, $Q(-L/2, t) =
Q(L/2,t)$, is given by
\begin{align}
  \displaystyle Q(z,t) = &\frac{2 \sigma_z A_g\sqrt{2\pi}}{L}
  \sum_{n=0}^{\infty}
  \bigg[\frac{\euler^{-q_n^2\sigma_z^2/2}}{c^2q_n^2+\kappa} \times
  \bigg( \cos(q_n z) \notag\\ &\times (1 - \cos
  (t\sqrt{c^2q_n^2+\kappa}))\bigg)\bigg],
\label{eq:qsol}
\end{align}
where $q_n = n \pi/L$.  This solution consists of a wavepacket of
standing waves weighted by a Gaussian distribution and a denominator
which comes from the phonon dispersion relation of
Eq.~\eqref{eq:quaddisp}.  The Gaussian distribution originates from
the Fourier transform of the Gaussian force in real space.  In this
case, the selection of $q$ is determined by the Fourier transform of
the driving force $S(z,t)$.  For the Gaussian force in our model, the
$q$ value can be selected for the region $0 < q < 1/\sigma_z$.  If the
maximum $q = 1/\sigma_z$ is smaller than $q_{\rm c} =
\sqrt{\kappa}/c$, then $Q(z,t)$ is localized.  If $1/\sigma_z$ is
larger than $q_{\rm c}$, then $Q(z,t)$ is divided into two
contributions: $0 < q < q_{\rm c}$ and $q_c \le q < 1/\sigma_z$, in
which the former $q$ value gives the localized wave and the latter
part gives the propagating wave.  We can then define a critical
parameter $\sigma_{z {\rm c}} = 1/q_c$ to explain the localization or
propagation of the coherent phonons obtained from the Klein-Gordon
equation.

For the (11,0) tube, we have a critical parameter $\sigma_{z {\rm c}}
= (2.545 / \sqrt{3147.22}) \unitnm = 0.045 \unitnm$.  Since in our
simulation we already used $\sigma_z = 0.9\unitnm$ which is much
larger than $\sigma_{z {\rm c}}$, it is then expected that the
coherent phonon is sufficiently localized.  To emphasize this fact, we
show two different cases of Klein-Gordon waves in
Fig.~\ref{fig:gaussdis} for $\sigma_z = 0.9\unitnm$ and $\sigma_z =
0.03\unitnm$.  Figure ~\ref{fig:gaussdis}(a) shows the two forces with
different $\sigma_z$ values, while Figs.~\ref{fig:gaussdis}(b) and (c)
shows the corresponding coherent phonon amplitudes that are generated.
It can be seen that we obtain localized (propagating) waves by using
$\sigma_z > \sigma_{z {\rm c}}$ ($\sigma_z < \sigma_{z {\rm c}}$).
Intuitively, we can understand from Fig.~\ref{fig:gaussdis}(c) that a
faster appearance of an amplitude propagating along the $z$ direction
can be obtained when $\sigma_z$ becomes much smaller than $\sigma_{z
  {\rm c}}$ although some parts of $Q(z,t)$ remain localized
(contribution from $0 < q < q_{\rm c}$).  The propagating wave
components in Fig.~\ref{fig:gaussdis}(c) travel with a velocity
$\sqrt{\kappa} / q$, where $q$ in this case is related to $\sigma_z$
directly by $q = 1/\sigma_z$, thus giving a speed of
$\sqrt{3147.22\unitps^{-2}} \times 0.03\unitnm = 1.68\unitnm/\unitps$.
In contrast, in the case of $\sigma_z$ much larger than $\sigma_{z
  {\rm c}}$ [e.g. Fig.~\ref{fig:gaussdis}(b)], we cannot see any
amplitudes along the $z$ direction except in a limited region where
the force exists, i.e. the propagating wave components cannot be
observed.  Indeed, the actual RBM dispersion is a bit flatter than the
approximation from the Klein-Gordon wave dispersion (see
Fig.~\ref{fig:RBMdisp}).  This means that the modes are localized even
more.  Therefore, in our case of $\sigma_z = 0.9\unitnm$, each
excitonic force will not interfere with neighboring force sites
separated by distance $d$, which indicates that the average amplitude
$A(t)$ in Fig.~\ref{fig:Aintegrated} is not affected by a random
separation between every excitonic force.  In general, we may say that
the localized vibration is a characteristic of the optical phonon
propagation driven by a localized force because the wavepacket is
dominated by $q\approx 0$ phonons, while the contribution of the group
velocity comes from $q \ge q_{\rm c}$.  This optical phonon feature
differs from that of the acoustic phonon feature whose solution is
expressed in terms of traveling waves.~\cite{sanders01}

We then calculate the optical absorption spectra as a function of time
using the calculated $Q(z,t)$.  It is expected that the inhomogeneous
coherent phonon oscillations induce a macroscopic atomic displacement
which modifies the transfer integral and thus modulates the energy
gap.  We calculate the absorption coefficient $\alpha (E_{\rm L}, t)$,
where $E_{\rm L}$ is the laser excitation energy, by evaluating it in
the dipole approximation using Fermi's golden rule.  The absorption
coefficient at a photon energy $E_{\rm L}$ obtained by including
exciton effects is given by~\cite{elliot57-absexc,konabe09-auger}
\begin{align} \label{Absorption coefficient} \alpha(E_{\rm L},t) =&
  \frac{8 e^2}{E_{\rm L} R m_0 c_0} \sum_{\mu k} \ \arrowvert
  \exop^\mu
  \arrowvert^2 \notag\\
  &\times \delta f^\mu (k,t) \ \delta \left( E_{ii}(t)- E_{\rm L}
  \right),
\end{align}
where $\exop^\mu$ is the exciton-photon matrix element within the
dipole approximation, corresponding to the transition between the
initial and final state on the $\mu$-th cutting line, $R$ is the tube
radius, $m_0$ is the electron mass, and $c_0$ is the speed of light.
The exciton energy $E_{ii}$ is now time-dependent because of the
change in transfer integral due to coherent RBM phonon vibrations
$A(t)$.

\begin{figure}[t]
  \centering \includegraphics[clip,width=6.5cm]{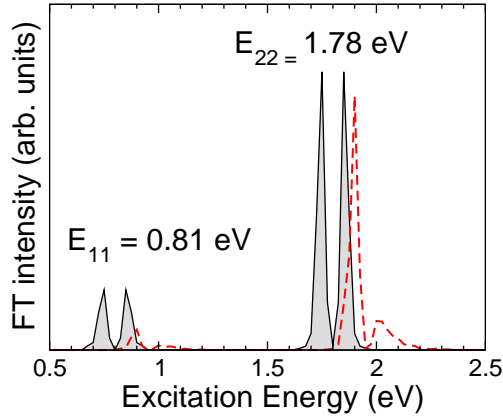}
  \caption{\label{fig:cpspectra} (Color online) Fourier transform
    intensity of the time-dependent absorption coefficient for the
    coherent RBM phonon of a (11,0) nanotube as a function of
    excitation energies $E_{\rm L}$.  The solid line represents the
    coherent phonon spectra which include excitonic effects, showing a
    symmetric double-peaked line shape at each transition energy
    $E_{ii}$.  The dashed line represents the coherent phonon spectra
    without excitonic effects, in which asymmetric line shapes were
    obtained previously.~\cite{sanders09-cp} }
\end{figure}

Since the bandgap is inversely proportional to the diameter
oscillation (or to the coherent RBM amplitudes), the time-dependent
absorption $\alpha(E_{\rm L}, t)$ has the same oscillating feature as
the average amplitude $A(t)$.  However, exciton effects acting on the
absorption spectrum will modify the shape of the absorption spectra
compared to that obtained without inclusion of the exciton effects.
We should then calculate the time-dependent absorption for a broad
range of excitation energies, for example, within the range of $0.5$
to $2.5\unitev$.  By performing a Fourier transformation numerically
over this energy range, we can obtain the RBM coherent phonon spectra
as shown in Fig.~\ref{fig:cpspectra}, which include $E_{11}$ and
$E_{22}$ for the $(11,0)$ tube that we consider.  The coherent phonon
spectra given in
Fig.~\ref{fig:cpspectra} show double-peaked structures as a function
of the excitation energies, either with or without including the
excitonic effects, as indicated by the solid and dashed lines in
Fig.~\ref{fig:cpspectra}, respectively.

The reason for the presence of the double-peak features (either
symmetric or asymmetric) in the excitation-dependent coherent phonon
intensity can be explained as follows.  The generation of coherent RBM
phonons modifies the electronic structure of SWNTs and thus it can be
detected as temporal oscillations in the transmittance of the probe
beam.  Since the RBM is an isotropic vibration of the nanotube lattice
in the radial direction, i.e. the diameter periodically oscillates at
the RBM frequency, this makes the band gap $E_g$ also oscillate at the
same frequency.  As a result, interband transition energies oscillate
in time, leading to ultrafast modulations of the absorption
coefficients at the RBM frequency, which is also equivalent to the
oscillations in the probe transmittance, and thus correspondingly, the
excitation energy dependence of the coherent phonon intensity shows a
derivative-like behavior.  More explicitly, the effect on the
absorption $\alpha$ for small changes in the gap can be modeled
by~\cite{kim13-cp}
\begin{equation}
  \alpha(E_{\rm L} - E_g) \approx \alpha(E_{\rm L}-E_g^0) - \frac{\partial
    \alpha(E_{\rm L}-E_g^0)}{\partial E_{\rm L}}  \delta E_g + \ldots,
\end{equation}
which gives
\begin{equation}
  \ \Delta \alpha \approx - \frac{\partial
    \alpha(E_{\rm L}-E_g^0)}{\partial E_{\rm L}}  \delta E_g ,
\end{equation}
where $E_g$ is assumed to be time-dependent, and $\delta E_g$ here
corresponds to a small change in the bandgap.  Since the coherent
phonon intensity is obtained by taking the Fourier transform (power
spectrum) of the differential transmission, the coherent phonon
intensity is thus proportional to the square of the derivative of the
absorption coefficient.

The excitonic absorption coefficient basically has a symmetric
lineshape with a single peak.~\cite{spa04-exc} Therefore, the
derivative of the excitonic absorption coefficient will give a
symmetric double-peak feature, in contrast to the asymmetric lineshape
expected from the 1D van Hove singularity (joint density of states).
Here the use of the Klein-Gordon equation which gives nonhomogeneous
macroscopic atomic displacements is then also justified by obtaining
the symmetric line shape for the coherent phonon spectra.  On the
other hand, in the free carrier model without the excitonic effects,
we see an asymmetric double-peaked structure at each transition with
the stronger peak at lower energy and the weaker peak at higher
energy, which originate from the derivative of the asymmetric
lineshape of the absorption coefficient.  Moreover it has also been
noted in some earlier works that the transition energy was shifted
upward by several hundred meV.~\cite{spa04-exc,kim13-cp}

As a final remark, we would like to mention that considering the
localized excitons in this work might be just one possibility that
gives the symmetric peak of the absorption spectrum because the origin
of the symmetric absorption lineshape is basically from the presence
of discrete energy levels of excitons in carbon nanotubes. In this
sense, if there are other configurations of excitons in carbon
nanotubes, which are not localized, such cases might also give rise to
the symmetric absorption lineshape. This can be an open issue for
future studies. However, we expect that as an initial condition of the
system after the excitation by the pump pulse, the excitons should be
localized with a certain average separation.~\cite{chang04-locexc,
  hirori06-locexc, hogele08-locexc}

\section{Conclusion}
\label{sec:conclude}

We have shown that excitonic effects modify the coherent phonon
amplitudes in SWNTs as described by the Klein-Gordon equation.  The
localized exciton wavefunctions result in an almost periodic and
localized driving force in space.  Although the exciton effects make
the amplitudes inhomogeneous, these amplitudes might be difficult to
observe in experiments where the long wavelength of the probe pulse
averages over the sample.  We then defined a spatial average of the
amplitudes that matches the experimental results.  Such an
interpretation becomes necessary and fundamental since we may say that
the pump-probe experiments on coherent phonons could not measure the
``real'' coherent phonon amplitudes of SWNTs.  What is measured in the
experiments is actually the average of the amplitudes.  Nevertheless,
using the present treatment we have been able to simulate the
experimental observation of a symmetric double-peaked structure as is
observed in the coherent phonon spectra as a function of excitation
energy.

\section*{Acknowledgments}
Tohoku University authors acknowledge financial support from JSPS,
Monbukagakusho Scholarship, and MEXT Grant No. 25286005.
E.R. contributed to this work during an undergraduate exchange program
(NanoJapan) funded by the PIRE project of NSF-OISE Grant No. 0968405.
Florida University authors acknowledge NSF-DMR Grant No. 1105437 and
OISE-0968405.  M.S.D. acknowledges NSF-DMR Grant No. 1004147.  We are
all grateful to Prof. J. Kono (Rice University) and his co-workers for
fruitful discussions which stimulated this work.

\bibliographystyle{aip}
%\bibliography{/home/students/nugraha/bib/nugraha}

\begin{thebibliography}{10}

\bibitem{c471}
R.~Saito, M.~Fujita, G.~Dresselhaus, and M.~S. Dresselhaus, Phys. Rev. B {\bf
  46}, 1804--1811 (1992).

\bibitem{gambetta06-cp}
A.~Gambetta, C.~Manzoni, E.~Menna, M.~Meneghetti, G.~Cerullo, G.~Lanzani,
  S.~Tretiak, A.~Piryatinski, A.~Saxena, R.~L. Martin, and A.~R. Bishop, Nat.
  Phys. {\bf 2}, 515--520 (2006).

\bibitem{lim06-cpexp}
Y.~S Lim, K.~J. Yee, J.~H. Kim, E.~H. Haroz, J.~Shaver, J.~Kono, S.~K. Doorn,
  R.~H. Hauge, and R.~E. Smalley, Nano Lett. {\bf 6}, 2696--2700 (2006).

\bibitem{kato08-cpaligned}
K.~Kato, K.~Ishioka, M.~Kitajima, J.~Tang, R.~Saito, and H.~Petek, Nano Lett.
  {\bf 8}, 3102--3108 (2008).

\bibitem{kim09-cpprl}
J.-H. Kim, K.-J. Han, N.-J. Kim, K.-J. Yee, Y.-S. Lim, G.~D. Sanders, C.~J.
  Stanton, L.~G. Booshehri, E.~H. H\'aroz, and J.~Kono, Phys. Rev. Lett. {\bf
  102}, 037402 (2009).

\bibitem{makino09-cpdoping}
K.~Makino, A.~Hirano, K.~Shiraki, Y.~Maeda, and M.~Hase, Phys. Rev. B {\bf 80},
  245428 (2009).

\bibitem{kilina07-adv}
S.~Kilina and S.~Tretiak, Adv. Func. Mat. {\bf 17}, 3405--3420 (2007).

\bibitem{sanders09-cp}
G.~D. Sanders, C.~J. Stanton, J.-H. Kim, K.-J. Yee, Y.-S. Lim, E.~H. H\'aroz,
  L.~G. Booshehri, J.~Kono, and R.~Saito, Phys. Rev. B {\bf 79}, 205434 (2009).

\bibitem{nugraha11-cp}
A.~R.~T. Nugraha, G.~D. Sanders, K.~Sato, C.~J. Stanton, M.~S. Dresselhaus, and
  R.~Saito, Phys. Rev. B {\bf 84}, 174302 (2011).

\bibitem{ando97-exc}
T.~Ando, J. Phys. Soc. Jpn. {\bf 66}, 1066--1073 (1997).

\bibitem{spa04-exc}
C.~D. Spataru, S.~Ismail-Beigi, L.~X. Benedict, and S.~G. Louie, Phys. Rev.
  Lett. {\bf 92}, 077402 (2004).

\bibitem{wang05-exp}
F.~Wang, G.~Dukovic, L.~E. Brus, and T.~F. Heinz, Science {\bf 308}, 838--841
  (2005).

\bibitem{jiang07-exc}
J.~Jiang, R.~Saito, Ge.~G. Samsonidze, A.~Jorio, S.~G. Chou, G.~Dresselhaus,
  and M.~S. Dresselhaus, Phys. Rev. B {\bf 75}, 035407 (2007).

\bibitem{jiang07-exphop}
J.~Jiang, R.~Saito, K.~Sato, J.~S. Park, Ge.~G. Samsonidze, A.~Jorio,
  G.~Dresselhaus, and M.~S. Dresselhaus, Phys. Rev. B {\bf 75}, 035405 (2007).

\bibitem{carsten10-exclocal}
Carsten Georgi, Alexander~A. Green, Mark~C. Hersam, and Achim Hartschuh, ACS
  Nano {\bf 4}, 5914--5920 (2010).

\bibitem{kamm07-biexciton}
D.~Kammerlander, D.~Prezzi, G.~Goldoni, E.~Molinari, and U.~Hohenester, Phys.
  Rev. Lett. {\bf 99}, 126806 (2007).

\bibitem{matsuda08-exciton}
K.~Matsuda, T.~Inoue, Y.~Murakami, S.~Maruyama, and Y.~Kanemitsu, Phys. Rev. B
  {\bf 77}, 033406 (2008).

\bibitem{stanton94-cpmethod}
A.~V. Kuznetsov and C.~J. Stanton, Phys. Rev. Lett. {\bf 73}, 3243--3246
  (1994).

\bibitem{merlin97-cp}
R.~Merlin, Solid State Commun. {\bf 102}, 207--220 (1997).

\bibitem{jishi93-phonon}
R.A. Jishi, L.~Venkataraman, M.S. Dresselhaus, and G.~Dresselhaus, Chem. Phys.
  Lett. {\bf 209}, 77--82 (1993).

\bibitem{maul02-phonon}
J.~Maultzsch, S.~Reich, C.~Thomsen, E.~Dobard«öi««, I.~Milo«Þevi««, and
  M.~Damnjanovi««, Solid State Comm. {\bf 121}, 471--474 (2002).

\bibitem{dubay03-phonon}
O.~Dubay and G.~Kresse, Phys. Rev. B {\bf 67}, 035401 (2003).

\bibitem{konabe09-auger}
S.~Konabe, T.~Yamamoto, and K.~Watanabe, Appl. Phys. Express {\bf 2}, 092202
  (2009).

\bibitem{xie12-diff}
J.~Xie, T.~Inaba, R.~Sugiyama, and Y.~Homma, Phys. Rev. B {\bf 85}, 085434 (Feb
  2012).

\bibitem{maru10-diffusion}
S.~Moritsubo, T.~Murai, T.~Shimada, Y.~Murakami, S.~Chiashi, S.~Maruyama, and
  Y.~K. Kato, Phys. Rev. Lett. {\bf 104}, 247402 (2010).

\bibitem{hagen05-decay}
A.~Hagen, M.~Steiner, M.~B. Raschke, C.~Lienau, T.~Hertel, H.~Qian, A.~J.
  Meixner, and A.~Hartschuh, Phys. Rev. Lett. {\bf 95}, 197401 (Oct 2005).

\bibitem{onoda11}
N.~Onoda, S.~Konabe, T.~Yamamoto, and K.~Watanabe, Phys. Status Solidi C {\bf
  8}, 570--572 (2011).

\bibitem{sanders01}
G.~D. Sanders, C.~J. Stanton, and Chang Sub~Kim, Phys. Rev. B {\bf 64}, 235316
  (2001).

\bibitem{elliot57-absexc}
R.~J. Elliott, Phys. Rev. {\bf 108}, 1384--1389 (1957).

\bibitem{kim13-cp}
J.-H. Kim, A.R.T. Nugraha, L.G. Booshehri, E.H. H«¡roz, K.~Sato, G.D. Sanders,
  K.-J. Yee, Y.-S. Lim, C.J. Stanton, R.~Saito, and J.~Kono, Chem. Phys. {\bf
  413}, 55--80 (2013).

\bibitem{chang04-locexc}
E.~Chang, G.~Bussi, A.~Ruini, and E.~Molinari, Phys. Rev. Lett. {\bf 92},
  196401 (2004).

\bibitem{hirori06-locexc}
H.~Hirori, K.~Matsuda, Y.~Miyauchi, S.~Maruyama, and Y.~Kanemitsu, Phys. Rev.
  Lett. {\bf 97}, 257401 (2006).

\bibitem{hogele08-locexc} A.~H\"ogele, C.~Galland, M.~Winger, and
  A.~Imamo\u{g}lu, Phys. Rev. Lett. {\bf 100}, 217401 (2008).

\end{thebibliography}

\end{document}